\documentclass[twocolumn,floatfix,prl]{revtex4}%
\usepackage[dvipdfmx]{graphicx}%
\usepackage{amsmath}%
\usepackage{amsfonts}
\usepackage{amssymb}
\usepackage{hhline}
\usepackage{bm}
\usepackage{mathrsfs}
\usepackage{color}

\def\s{{\sigma}}
\def\e{{\epsilon}}
\def\k{{ {\bm k} }}
\def\p{{ {\bm p} }}
\def\q{{ {\bm q} }}

\def\0{{ {\bm 0} }}
\def\w{{\omega}}
\def\a{{\alpha}}
\def\b{{\beta}}

\allowdisplaybreaks[4]

\begin{document}
\title{Hexadecapole fluctuation mechanism for $s$-wave 
heavy fermion superconductor CeCu$_2$Si$_2$:
Interplay between intra- and inter-orbital Cooper pairs
}
\author{
Rina Tazai, and
Hiroshi Kontani
}

\date{\today }

\begin{abstract}
In heavy-fermion superconductors, it is widely believed that 
the superconducting gap function has sign-reversal due to the 
strong electron correlation.
However, recently discovered fully-gapped $s$-wave 
superconductivity in CeCu$_2$Si$_2$
has clarified that strong attractive pairing interaction can appear
even in heavy-fermion systems.
To understand the origin of attractive force,
we develop the multipole fluctuation theory 
by focusing on the inter-multipole many-body interaction 
called the vertex corrections.
By analyzing the periodic Anderson model for CeCu$_2$Si$_2$,
we find that hexadecapole fluctuations mediate strong attractive 
pairing interaction.
Therefore, fully-gapped $s$-wave superconductivity is driven 
by pure on-site Coulomb repulsion,
without introducing electron-phonon interactions.
The present theory of superconductivity
will be useful to understand rich variety of the superconducting 
states in heavy fermion systems.

\end{abstract}

\address{
Department of Physics, Nagoya University,
Furo-cho, Nagoya 464-8602, Japan.
}
 

\sloppy

\maketitle


Heavy fermion (HF) systems exhibit wide variety of
unconventional superconductivities
\cite{Stewart,Peid,Maple}.
For example, antiferro- and ferro-
magnetic dipole (rank 1) fluctuations mediate
interesting pairing states, such as
$d$-wave singlet pairing in Ce$M$In$_5$ ($M$=Rh,Co,Ir) 
\cite{Izawa-115}
and triplet pairing in UCoGe \cite{Ishida-UCoGe}.
Since the magnetic dipole fluctuations mediate
repulsive pairing interaction,
the superconducting gap function inevitably has sign-reversal
\cite{Moriya,Yamada,Kontani-rev,Scalapino,Takimoto-SC}.
However, there are many pairing states in HF systems that
cannot be understood based by the rank 1 fluctuations mechanism.
In HF systems, it is noteworthy that 
higher-rank ($r\ge2$) multipole operators are also active 
thanks to the strong spin-orbit interaction (SOI),
and therefore rich multipole physics emerges.
Although higher-rank multipole fluctuations in principle 
cause exotic pairing states,
theoretical studies have not been performed enough.



CeCu$_2$Si$_2$ is a famous HF superconductor
near the magnetic criticality
\cite{Ste-122,Yuan-122,Ishida,Hol-122},
and recently reported fully-gapped structure in 
CeCu$_2$Si$_2$ attract considerable attention
\cite{Kit1-122,Kit2-122,Yama-122,Steglich}.
The absence of nodes is confirmed by the measurements of
the specific heat, thermal conductivity and penetration depth
for $T\ll T_{\rm c}$.
In addition, robustness of $T_{\rm c}$ against randomness
strongly indicates the plain $s$-wave state without any sign-reversal
\cite{Yama-122}.
Theoretically, magnetic multipole (MM) ($r=1,3,5$) fluctuations
will cause sign-reversing pairing states
\cite{Ikeda-122}.
Therefore, electric multipole (EM) ($r=2,4$) fluctuations 
that give attractive pairing interaction 
would be important in CeCu$_2$Si$_2$,
whereas the microscopic origin of EM fluctuations is unknown.

The minimum theoretical model of CeCu$_2$Si$_2$ 
is the four-orbital ($J_z=\pm5/2,\pm3/2$)
periodic Anderson model (PAM) with on-site Coulomb interaction.
However, if we apply the random-phase-approximation (RPA)
to this model, none of EM fluctuations develop.
This negative result indicates the significance of the 
vertex corrections (VCs), which represent the 
many-body effects beyond the RPA.
Recently, it was revealed that 
higher-rank multipole fluctuations develop cooperatively
due to the Aslamazov-Larkin (AL) type $\chi$-VC,
which is the VC for the susceptibility,
in the study of multipole order in CeB$_6$
\cite{Tazai-CeB6}.
Physically, the AL-VC gives strong interference
between EM and MM fluctuations.
Also, the attractive pairing interaction
(such as phonon-mediated interaction)
is strongly magnified by the $U$-VC, which is the VC 
for the electron-boson coupling in the gap equation
\cite{Tazai-PRB2018}.
Considering these VCs properly,
mysterious plain $s$-wave superconductivity in CeCu$_2$Si$_2$ 
may be understood in terms of the EM fluctuation mechanism,
even if the $e$-ph interaction is absent.

In this paper, we develop a theory of 
multipole fluctuation mediated superconductivity 
in HF systems based on the multiorbital PAM.
Due to the AL-VC for susceptibility ($\chi$-VC),
strong quadrupole and hexadecapole fluctuations develop
even in the absence of $e$-ph interaction.
In CeCu$_2$Si$_2$, the hexadecapole fluctuations mediate
strong attractive pairing interaction, and it is magnified 
by the AL-VC for the electron-boson coupling ($U$-VC)
in the gap equation.
Thus, fully-gapped $s$-wave state is caused by the 
hexadecapole fluctuations 
against strong on-site repulsive Coulomb interaction.
The present pairing mechanism may be significant 
to understand various HF superconductors.

Now, we introduce a two-dimensional $J=5/2$ PAM for CeCu$_2$Si$_2$.
According to the LDA+DMFT study \cite{LDADMFT_multiporbital},
the following $f$-electron states 
in $J_{z}$-basis are important near the Fermi level:
$|f_{1},\Sigma \rangle=|\mp \frac{5}{2}\rangle$ and 
$|f_{2}, \Sigma \rangle=-|\pm \frac{3}{2}\rangle$,
where $\Sigma=\pm$ denotes pseudo-spin 
\cite{Tazai-PRB2018,RIXS}.
The kinetic term of the $\Gamma_7^{(1)}$-$\Gamma_7^{(2)}$ quartet PAM is given by
\begin{eqnarray}
\hat{H}_{0}=\sum_{\k\sigma}\epsilon_{\k}c^{\dagger}_{\k\sigma}c_{\k\sigma}+\sum_{\k l\sigma}E_{\k l}f^{\dagger}_{\k l\sigma}f_{\k l\sigma}+
\left(V^{*}_{\k l\sigma}f^{\dagger}_{\k l\sigma}c_{\k\sigma}+{\rm h.c}\right) ,
\nonumber 
\end{eqnarray}
where $c^{\dagger}_{\k \sigma}$ ($f^{\dagger}_{\k l \sigma}$) 
is a creation operator for $s\ (f_l)$-electron
with momentum $\k$.
Here, we put $\Sigma=\sigma$ since the 
pseudo-spin is conserved in the present PAM
\cite{Tazai-PRB2018}.
We set $\epsilon_{\k}=2t_{ss}(\cos k_{x}+\cos k_{y})+\epsilon_{0}$
and $E_{\k l}=E^f_l-(-1)^l\delta E_{\k}$ ($l=1,2$).
Here, $\delta E_{\k}$ is given by small $f$-$f$ hopping integrals
($|\delta E_{\k}|<0.12|t_{ss}|$)
as we explain in the supplemental material (SM) A\cite{SM}.
$V_{\k l\sigma}$ is the $f$-$s$
hybridization term between the nearest sites, given as
$
V_{\k l \sigma}= (-1)^{l} t_{sf}^{l}(\sin k_{y} - i\sigma\sin k_{x} )
$
\cite{Tazai-PRB2018}.
To make the analysis simple, we set 
$E_{1}^f=E_{2}^f \equiv E^f$ and $t_{sf}^{1}=t_{sf}^{2} \equiv t_{sf}$.
Then, the relation $D_{1}(\e)=D_{2}(\e)$ holds, where 
$D_l(\e)$ is the density of states (DOS) of $f_l$-electrons.
This is consistent with the relation 
$D_{1}(0)\approx D_{2}(0)$ given by LDA+DMFT study of CeCu$_2$Si$_2$
\cite{LDADMFT_multiporbital,comment}.
In the following numerical study,
we set $t_{ss}=-1.0$, $E^f=0.1$,
$\epsilon_{0}=3.0$, $t_{sf}=0.62$, 
temperature $T=0.045$ and the chemical potential $\mu=-0.143$.
Then, $f(s)$-electron number is $n_{f}=0.9$ ($n_{s}=0.3$).

\begin{figure}[htb]
\includegraphics[width=.96\linewidth]{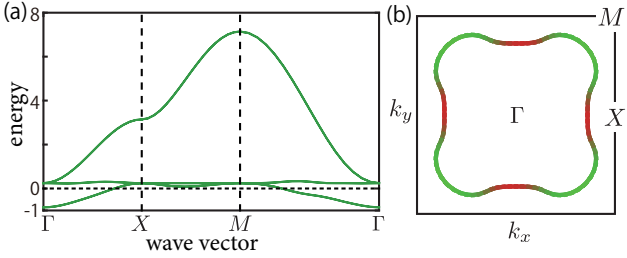}
\caption{
(a) Band dispersion and (b) the FS of the present PAM with 
$\delta E_\k$ (model A).
Red (green) color corresponds to the weight of 
$f_{1}(f_{2})$-orbital, $w_1(\k)$ ($w_2(\k)$).
The ratio $w_1(\k)/(w_1(\k)+w_2(\k))$ 
on the FS ranges about $0.1-0.9$.
}
\label{fig:band}
\end{figure}

In Fig.\ref{fig:band} (a), we show the band structure.
$\epsilon=0$ corresponds to the Fermi level.
The total band width is $W_{D}\sim 10$ (in unit $|t_{ss}|=1$),
and $W_{D}\sim 10$eV in CeCu$_2$Si$_2$ \cite{Ikeda-122}.
The width of quasi-particle band (=the lowest band) is 
$W_{D}^{qp} \sim 1$.
The Fermi surface (FS) is shown in Fig.\ref{fig:band} (b).
The anisotropy of $f_{l}$-orbital weight on the FS
is introduced by $\delta E_{\k l}$,
which exists in real HF compounds.
We call the present PAM with orbital anisotropy the model A.
We will discuss later that the orbital anisotropy is favorable
for the $s$-wave state. 

We introduce the interaction term
$\hat{H}_{U}=u \hat{H}^{0}_{U}$.
Here, $\hat{H}^{0}_{U}=\frac14 \sum_{LL'MM'} U^0_{L,L';M,M'}
f_{L}^\dagger f_{L'} f_{M} f_{M'}^\dagger$,
where $L=(l,\sigma)$ and $M=(m,\rho)$.
$\hat{U}^0$ is the $16\times16$
normalized Coulomb interaction,
of which the maximum element is unity
\cite{Tazai-PRB2018}.
The pseudo-spin is conserved in $\hat{H}_{U}$.

The present model belongs to $D_{4h}$ point group.
The active irreducible representation (IR) are 
$\Gamma^+=A_{1}^{+},A_{2}^{+},E^{+}$ and 
$\Gamma^-=A_{1}^{-},A_{2}^{-},E^{-}$ 
\cite{Tazai-PRB2018}.
In TABLE \ref{tab:multipole}, we show 
the active EM operators
and their approximate pseudo-spin representations.
The $4\times4$ matrix form of each multipole operator ${\hat Q}$ 
is shown in the SM B \cite{SM}.

\begin{table}[htb]
  \begin{tabular}{|c|c|c|c|} \hline
     IR ($\Gamma$) & rank (k) & operator ($\hat{Q}$) & matrix \\ \hhline{|=|=|=|=|}
     & $0$  & $\hat{C}$ & $\hat{\sigma}^{0}\hat{\tau}^{0}$ \\ \cline{2-4}
     $A_{1}^{+}$& $2$  & $\hat{O}_{20}$ &  $\hat{\sigma}^{0}(3\hat{\tau}^{z}+2\hat{\tau}^0)$   \\ \cline{2-4}
      & $4$  & $\hat{H}_{0}$ & $\hat{\sigma}^{0}(-2.2\hat{\tau}^x+2\hat{\tau}^{z}-\hat{\tau}^{0})$ \\ \hline
    $A_{2}^{+}$ & $4$  & $\hat{H}_{z}$ & $\hat{\sigma}^{z} \hat{\tau}^{y}$  \\   \hline
    $E^{+}$ & $2$  & $\hat{O}_{yz},\hat{O}_{zx}$ &  $\hat{\sigma}^{x} \hat{\tau}^{y}$, $\hat{\sigma}^{y} \hat{\tau}^{y}$ \\  \hline
    \end{tabular}
    \caption{Simple expressions of active EM operators
in the $\Gamma_7^{(1)}$-$\Gamma_7^{(2)}$ quartet model.
}
    \label{tab:multipole}
\end{table}

\begin{figure}[htb]
\includegraphics[width=.7\linewidth]{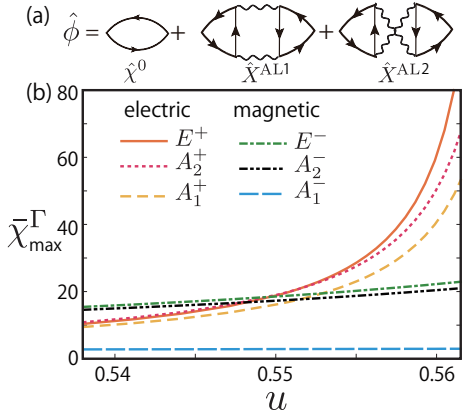}
\caption{
(a) Diagram of irreducible susceptibility with AL-VCs.
(b) Obtained susceptibility for each IR. 
EM susceptibilities ($\Gamma=E^{+},A^{+}_{2},A^{+}_{1}$) 
develop due to the AL-VC.
}
\label{fig:chi}
\end{figure}

From now on, we calculate the $f$-electron susceptibilities.
The bare irreducible susceptibility 
is $\chi_{\a,\b}^{0}(q)= -T\sum_{k}G^{f}_{LM}(k+q)G^{f}_{M' L'}(k)$, 
where  $q\equiv (\q, \omega_{j})=(\q,2j\pi T)$, 
$\a\equiv (L,L')$ and $\b\equiv (M,M')$.
$\hat{G}^f$ is the $f$-electron Green function without self-energy
\cite{Tazai-PRB2018}.
To go beyond the RPA,
we calculate the AL term for $\chi$-VC, $\hat{X}^{\rm{AL}}$.
Its diagrammatic expression and analytic one
are respectively given in Fig.\ref{fig:chi} (a) and in the SM C \cite{SM}.
Since $\chi$-VC is important only for EM susceptibilities,
we project out the magnetic channel contribution of $\chi$-VC
\cite{Yamakawa-FeSe,Tazai-CeB6,Onari-SCVC,rina2,rina1,Tsuchiizu}.
We also drop the MT-type VC since its contribution is small
\cite{Yamakawa-FeSe,Tazai-CeB6,Onari-SCVC,rina2,rina1,Tsuchiizu}.
Then, the $f$-electron susceptibility 
in the $16\times16$ matrix form 
is given as
\begin{eqnarray}
\hat{\chi}(q)= {\hat \phi}(q)
({\hat 1}-u{\hat U}^0{\hat \phi}(q))^{-1},
\label{eq:rpa}
\end{eqnarray}
where $\hat{\phi}(q)=\hat{\chi}^{0}(q)+\hat{X}^{\rm{AL}}(q)$ 
is irreducible susceptibility.
To derive the multipole susceptibility, 
we solve the following eigenvalue equation
\begin{eqnarray}
\hat{\chi}(\q,0)\vec{w}^{\Gamma}_\q=\bar{\chi}^{\Gamma}_\q \vec{w}^{\Gamma}_\q,
\label{eqn:eigenequation}
\end{eqnarray} 
where 
$\vec{w}^{\Gamma}$ is the eigenvector 
that belongs to the IR $\Gamma$.
It is expressed as
$\vec{w}^{\Gamma} = \sum_{Q\in\Gamma}b_Q{\vec Q}$,
where $b_Q$ is real coefficient and 
$\vec{Q}$ is $16\times 1$ vector defined as
$(\vec Q)_\a \equiv (\hat Q)_{L,M}$ with $\a=(L,M)$.
Then, 
the largest eigenvalue $\bar{\chi}^{\Gamma}_\q$ 
gives the multipole susceptibility for the IR $\Gamma$.

In Fig.\ref{fig:chi} (b), we show the obtained 
$\bar{\chi}^{\Gamma}_{\rm max} \equiv \max_\q\{\bar{\chi}^\Gamma_\q\}$
for each $\Gamma$.
With increasing $u$, all the EM fluctuations 
strongly develop thanks to the AL-VC.
Thus, large EM susceptibilities originate 
from the interference of MM fluctuations,
as discussed in the study of multipole order in CeB$_6$
\cite{Tazai-CeB6}.
For the EM susceptibilities,
the maximum position of $\bar{\chi}^{\Gamma}_\q$ for $\Gamma=A_1^+$
is $\q\approx(\pi,\pi)$,
whereas that for $\Gamma=A_2^+,E^+$ is $\q\approx(0,0)$.
For the MM susceptibilities,
the maximum position for $\Gamma=A_{2}^{-},E^{-}$
is $\q\approx(\pi/2,\pi/2)$.

In the next stage, 
we solve the linearized gap equation with $U$-VC
introduced in Ref. \cite{Tazai-PRB2018}:
%
\begin{eqnarray}
\lambda \Delta(k)=
\frac{\pi T}{(2\pi)^2}\sum_{\epsilon_{m}}
\oint_{\rm FS} \frac{d\p}{v_{\p}} 
\frac{\Delta(p)}{|\epsilon_{m}|} V^{\rm{sing}}_{kp} ,
\label{eqn:linear}
\end{eqnarray}
where $k=(\k,\e_n)=(\k,(2n+1)\pi T)$ and $p=(\p,\e_m)=(\p,(2m+1)\pi T)$.
$\Delta(k)$ is the gap function on the FS,
$\lambda$ is the eigenvalue, and
$v_{\p}$ is the Fermi velocity.
${V}^{\rm{sing}}_{kp}$ is the 
spin singlet paring interaction in band basis,
given by the unitary transformation of
$\hat{V}_{kp}^{\rm{sing}} \equiv \hat{V}^{\chi}_{kp}+\hat{V}^{U}_{kp}$.
Here,
\begin{eqnarray}
&&\hat{V}^{\chi}_{kp}= \{
u^2 \hat{\Lambda}_{kp} \hat{U}^0 \hat{\chi}(k-p) \hat{U}^0 \hat{\Lambda}'_{kp}
\}_{\uparrow \uparrow \downarrow \downarrow}
-\{ \quad \}_{\uparrow \downarrow \uparrow \downarrow},
\nonumber \\
&&\hat{V}^U_{kp} =u \{ \hat{\Lambda}_{kp} \hat{U}^0 \hat{\Lambda}'_{kp}
\}_{\uparrow \uparrow \downarrow \downarrow}
-\{ \quad \}_{\uparrow \downarrow \uparrow \downarrow},
\label{eqn:int}
\end{eqnarray}
where 
$V^{\chi}$ ($V^{U}$) gives the pairing interaction 
due to fluctuations (Coulomb repulsion).
$\hat{\Lambda}_{kp}$ and $(\hat{\Lambda}'_{kp})_{LL'MM'}\equiv
(\hat{\Lambda}_{kp})_{M'ML'L}$ are AL-type $U$-VCs
\cite{Tazai-PRB2018}.
The expression of $\hat{\Lambda}_{kp}$ is given in 
Ref. \cite{Tazai-PRB2018} and SM C \cite{SM}.

\begin{figure}[htb]
\includegraphics[width=.7\linewidth]{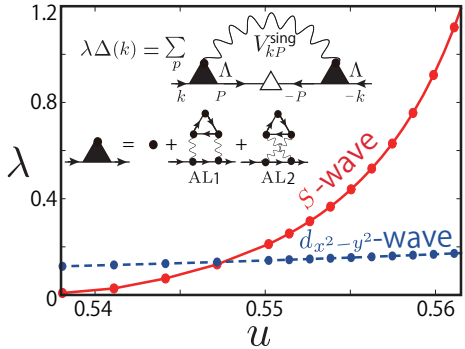}
\caption{
Obtained eigenvalue as the function of Coulomb interaction $u$. 
The $d_{x^2-y2}$-wave state is replaced with the 
fully gapped $s$-wave state for $u>0.55$.}
\label{fig:sc}
\end{figure}

The gap equation is schematically shown in the inset of Fig.\ref{fig:sc},
where the black triangle is the $U$-VC.
As explained in Ref. \cite{Tazai-PRB2018},
$|\hat{\Lambda}|^2\gg1$ for the electric channel
in the presence of MM fluctuations.
Therefore, the pairing interaction due to 
hexadecapole or quadrupole fluctuations is strongly enlarged 
by $|\hat{\Lambda}|^2\gg1$.
As shown in Fig.\ref{fig:sc}, when $u>0.55$, the $d_{x^2-y^2}$-wave 
state is replaced with the $s$-wave state mediated by the 
strong EM fluctuations in Fig. \ref{fig:chi} (b).
The obtained $s$-wave state is 
fully gapped without sign reversal,
consistently with experiments in CeCu$_2$Si$_2$.


Now, we discuss the origin of the $s$-wave superconductivity.
For this purpose, we decompose the susceptibility 
into the summation of $(Q,Q')$-channel as
\cite{Tazai-PRB2018,Tazai-CeB6}
%
\begin{eqnarray}
\hat{\chi}(q)= \sum_{QQ'}\hat{\chi}^{QQ'}(q), \ \
\hat{\chi}^{QQ'}(q)\equiv \chi^{QQ'}_q\vec{Q}(\vec{Q}')^{\dagger} ,
\label{eq:suscep}
\end{eqnarray} 
where $\chi^{QQ'}_q$ is s scalar multipole susceptibility.
Then, $\{\vec{Q}\}$ forms a complete non-orthogonal basis:
$(\vec{Q})^{\dagger}\vec{Q}'$ is unity for $Q=Q'$,
whereas it is zero when $Q$ and $Q'$ belong to different IR.
Note that $\bar{\chi}^\Gamma_q$ is the maximum eigenvalue of 
the Hermite matrix composed of $\chi^{QQ'}_q$ with $Q,Q'\in\Gamma$.

\begin{figure}[htb]
\includegraphics[width=.99\linewidth]{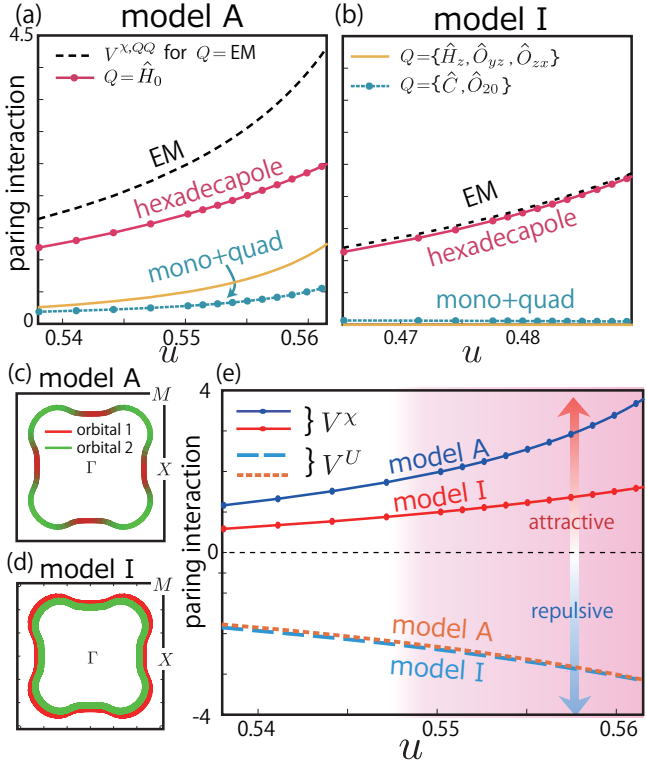}
\caption{
Obtained $V^{\chi,{QQ}}$ and $V^{\chi,{\rm EM}}$ 
due to the EM fluctuations
in (a) model A and (b) model I.
(c),(d) Orbital character on the FS.
(e) Obtained paring interactions $V^\chi$ and $V^U$.
For model I, the horizontal axis is shifted by $+0.073$.
}
\label{fig:hexa}
\end{figure}

Then, the $(Q,Q')$-fluctuation-induced paring interaction
in the band basis, $V^{{\chi},{QQ'}}_{kp}$, 
is given by the unitary formation of 
%
\begin{eqnarray}
\hat{V}^{\chi,{QQ'}}_{kp}\! 
= \{u^2 \hat{\Lambda}_{kp} \hat{U}^0 \hat{\chi}^{QQ'}(k-p) \hat{U}^0 \hat{\Lambda}'_{kp} 
\}_{\uparrow \uparrow \downarrow \downarrow}
-\{ \quad \}_{\uparrow \downarrow \uparrow \downarrow} .
\label{eqn:sc}
\end{eqnarray}
%
In Fig.\ref{fig:hexa} (a),
we show the EM-fluctuation-mediated interaction 
averaged on the FS,
$V^{\chi,{QQ'}} \equiv \int_{\rm FS} dkdp V^{\chi,{QQ'}}_{kp}/\int_{\rm FS} dkdp$,
for $Q=Q'$,
together with the total EM-fluctuation interaction
$V^{\chi,{\rm EM}}\equiv \sum_{QQ'}^{EM}V^{\chi,{QQ'}}$.
In the present model with $\delta E_\k$ (model A), 
the contribution from the hexadecapole ($H_{0}$) fluctuations
in the $A_{1}^+$ representation is the largest, while
other EM fluctuations are also important.
For comparison, we analyze the orbital isotropic model
with $\delta E_\k=0$, which we call the model I.
Surprisingly, in the model I, multipole fluctuations other than $H_{0}$
do not contribute to the $s$-wave pairing,
irrespective that all EM ($E^+,A_2^+,A_1^+$)
susceptibilities develop
similarly to Fig. \ref{fig:chi} (b) for model A.
The FS and its orbital character in each model
are shown in Figs.\ref{fig:hexa} (c) and (d).
In model I, the orbital weight is perfectly isotropic,
whereas the shape of FS is almost model-independent.



Figure \ref{fig:hexa} (e) shows the $s$-wave pairing interactions
$V^\chi$ and $V^U$ averaged over the FS.
Both $V^\chi$ and $|V^U|$ increase with $u$ in both models
due to large $U$-VC for the EM channel \cite{Tazai-PRB2018}.
In model I, the total pairing interaction $V^{\rm sing}=V^{\chi}+V^{U}$
is always negative (=repulsive), so the $d$-wave state appears.
In model A, in contrast, $V^{\rm sing}$ becomes positive with $u$
since not only $H_0$ fluctuations, but also other EM fluctuations 
contribute to the attractive pairing when $\delta E_\k\ne0$.
Therefore, the fully-gapped $s$-wave state is realized in model A.
As shown in Fig. \ref{fig:sc}, the eigenvalue $\lambda$ 
for the $s$-wave state is very large because of the 
retardation effect as we explain in the SM D \cite{SM}.
In fact, $V^{\rm sing}$ due to EM fluctuations
is attractive only for lower frequencies.

Finally, we discuss why all EM fluctuations 
contribute to the $s$-wave state in model A ($\delta E_\k\ne0$).
Since the relation $D_1(\e)\approx D_2(\e)$
holds even if $\delta E_\k$ exists,
the obtained EM- and MM-fluctuations are 
similar in both model A and I.
On the other hand, the ``inter-orbital pairing 
$\langle f_{\k 1 \uparrow} f_{-\k 2 \downarrow}\rangle$''
is suppressed in model A due to the $\k$-dependence 
of the orbital character on the FS.
The absence of inter-orbital pairing is favorable 
for $s$-wave state as we will discuss later.

One may expect that any EM fluctuations causes the 
attractive pairing interaction.
However, some elements of the EM susceptibility
${\hat \chi}^{QQ'}(\q,0)=\chi^{QQ'}_q\vec{Q}(\vec{Q'})^\dagger$
are negative except for $Q=Q'=C$, so
the cancellation of pairing interaction may occur.
(For example, 
$\{{\hat \chi}^{QQ}(\q)\}_{1\uparrow,2\downarrow;1\uparrow,2\downarrow}<0$
for $Q=O_{yz,zx}$.)

Now, we consider the gap equation 
when the pairing interaction is given as
$\hat{V}^{(\mu\nu)}= \bar{g} \vec{P}_{\mu\nu}(\vec{P}_{\mu\nu})^\dagger$,
where $\bar{g}>0$ and 
${\hat P}_{\mu\nu}\equiv\hat{\s}_{\mu}\hat{\tau}_{\nu}$ ($\mu,\nu=0,x,y,z$).
All the EM operators are given by 
(linear combination of) 
$\hat{P}_{\mu\nu}$ with $(\mu\nu)=(00),(0x),(0z),(xy),(yy),(zy)$.
The gap equation with BCS type cut-off $\w_c$ in the orbital basis is 
\begin{eqnarray}
\lambda \hat{\Delta} \approx T\sum_k \bar{g} \ ^t\!\hat{P}_{\mu\nu} 
\hat{G}^f(k) \hat{\Delta} \hat{G}^f(-k) \hat{P}_{\mu\nu}
\theta (\w_c-|\e_n|) .
\label{eq:hexa1}
\end{eqnarray}
%
As explained in Ref. \cite{Tazai-PRB2018},
$\hat{G}^f$ is expressed as
\begin{eqnarray}
G_{lm}^f(k)=G_{l}^0(k)\delta_{l,m}+(-1)^{l-m}
r_{\k}G_{l}^0(k)G_{m}^0(k)G^c(k),
\label{eqn:Green}
\end{eqnarray}
where $G^c(k)=(i\e_n-\e_\k-\sum_l r_{\k} G_l^0(k))^{-1}$ 
is the $c$-Green function,
$r_{\k}\equiv |V_{\k 1\uparrow}|^2$, and
$G_{l}^0(k)=(i\e_n-E_{\k l})^{-1}$ is the unhybridized $f$-Green function.
We neglect the first term in Eq. (\ref{eqn:Green}) 
since it does not give $-{\rm ln}T$ term in gap equation.
Therefore, in model I ($\delta E_\k=0$), the relation 
$\hat{G}\propto \hat{\s}_0(\hat{\tau}_0-\hat{\tau}_x)$ 
holds since $G_{1}^0=G_{2}^0$.
In model A with large $|\delta E_\k| \ (\gg|V_{\k 1 \uparrow}|^2)$,
the relation $\hat{G}\propto \hat{\s}_0\hat{\tau}_0$ holds approximately.

Here, we set
$\hat{G}\propto \hat{\sigma}_{0} (\hat{\tau}_{0}-a\hat{\tau}_{x})$
and $\hat{\Delta}\propto i \hat{\sigma}_{y}(\Delta_{0} 
\hat{\tau}_{0}+\Delta_{x} \hat{\tau}_{x})$:
$a=1$ $(a=0)$ corresponds to model I 
(model A with $|\delta E_\k|\gg |V_{\k 1 \uparrow}|^2$).
In this case,
\begin{eqnarray}
\hat{G} \hat{\Delta} \hat{G} &\propto& 
((1+a^2)\Delta_0-2a\Delta_x)i\hat{\s}_y\hat{\tau}_0
\nonumber \\
& &+((1+a^2)\Delta_x-2a\Delta_0)i\hat{\s}_y\hat{\tau}_x .
\end{eqnarray}
Then, the eigenvalue of the gap equation is
\begin{eqnarray}
&&\!\!\!\!\!\!
\lambda = g(1+a)^2 \ \ \ {\rm for} \ (\mu\nu)=(00),(0x), 
 \\
&&\!\!\!\!\!\!
\lambda = g(1-a)^2 \ \ \ {\rm for} \ (\mu\nu)=(0z),(xy),(yy),(zy),
\end{eqnarray}
where $g=\bar{g}D_1(0){\rm ln}(\w_c/T)$.

\begin{figure}[htb]
\includegraphics[width=.85\linewidth]{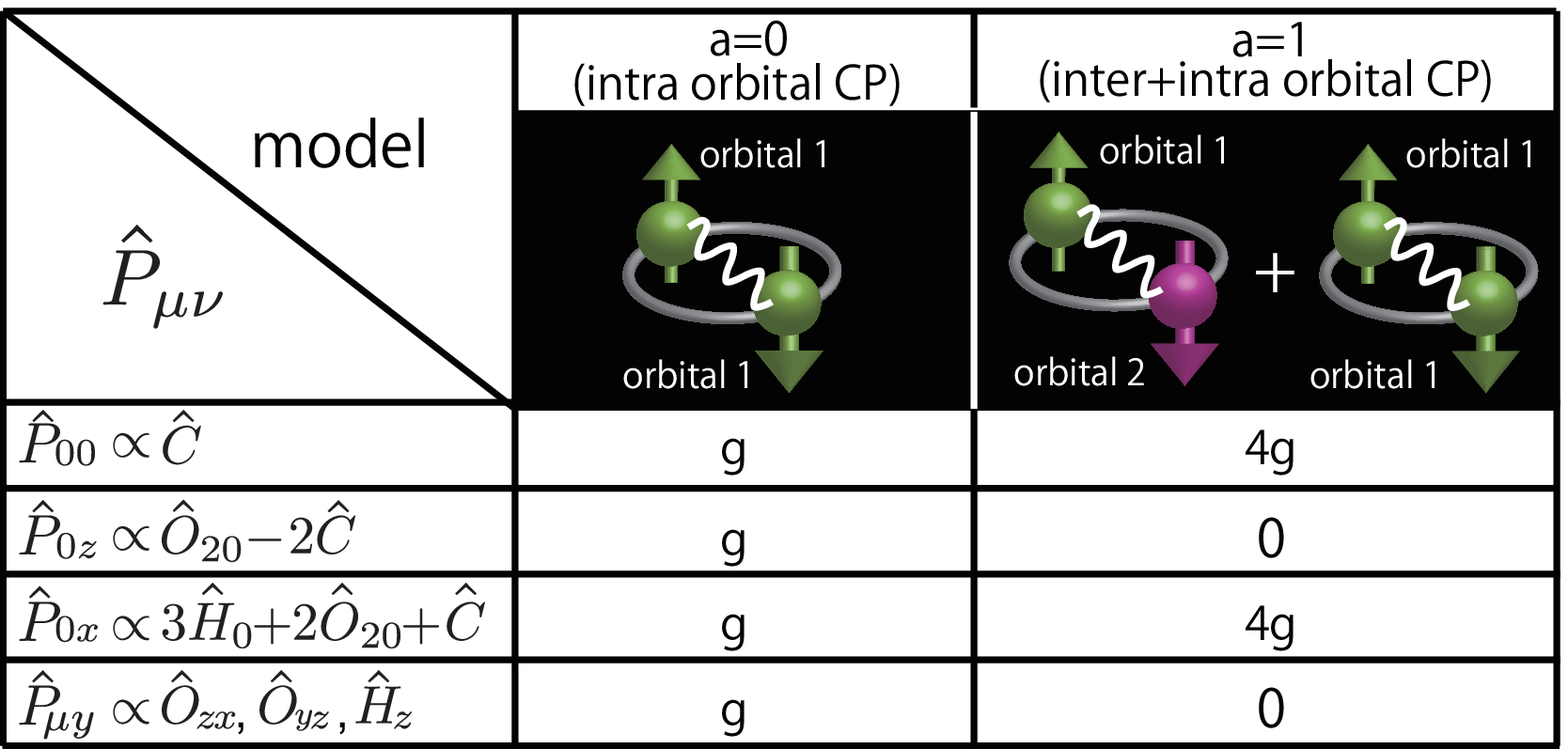}
\caption{
Eigenvalue $\lambda$ due to $\hat{P}_{\mu\nu}$ EM interaction
for $a=0$ (intra orbital Cooper pair (CP))
and $a=1$ (intra+inter orbital CP).
Except for $\hat{P}_{0x}$ and $\hat{C}=\hat{P}_{00}$, 
the EM fluctuations give
repulsive interaction for inter-orbital Cooper pair.
}
\label{fig:table}
\end{figure}

In Fig.\ref{fig:table}, 
we summarize the eigenvalue $\lambda$ 
for each EM pairing interaction,
in the case of $a=0$ (intra orbital Cooper pair)
and $a=1$ (intra+inter orbital Cooper pair).
We note that $\hat{P}_{0z}\propto \hat{O}_{20}-2 \hat{C}$
and $\hat{P}_{0x}\propto -3\hat{H}_0+2\hat{O}_{20}+\hat{C}$.
In case of $a=0$, all EM fluctuations contribute to the pairing.
In case of $a=1$, however, only $\hat{P}_{0x}$ and $C$ channels
contribute to the pairing.
In the present PAM,
charge ($C$) fluctuations are small, 
so they do not contribute to the pairing.
Since $\hat{P}_{0x}$ is included only in $H_0$ hexadecapole,
the $H_0$ fluctuations give dominant $s$-wave pairing interaction.
To summarize, the pairing interaction increases 
if the inter orbital Cooper pairs are killed by finite $|\delta E_\k|$,
so the numerical results in Fig.\ref{fig:hexa} are well understood.

In summary,
we studied the multipole fluctuation mediated superconductivity 
in HF systems based on the $\Gamma_7^{(1)}$-$\Gamma_7^{(2)}$ quartet PAM.
Due to the AL-type $\chi$-VC,
strong quadrupole and hexadecapole fluctuations develop,
and the resultant attractive interaction 
is enlarged by the $U$-VC in the gap equation.
In CeCu$_2$Si$_2$, $H_0$ hexadecapole fluctuations mediate
strong attractive pairing interaction.
The $s$-wave state is further stabilized by 
introducing small $\delta E_{\k l}$,
by which the inter-orbital Cooper pairs are killed.
Moreover, if we introduce the $e$-ph interaction,
both $U$-VCs ($\chi$-VC and $U$-VC) and $e$-ph interaction 
would enlarge $s$-wave $T_{\rm c}$ cooperatively
\cite{Tazai-PRB2018,Razaf,Ohkawa,Nagaoka,Kontani-RPA}.
The present pairing mechanism may be significant 
to understand various HF superconductors.

There are many important future issues, 
such as the self-energy effect
\cite{Haule,LDADMFT_multiporbital,KotGeo-DMFT,KotVol-DMFT,Held,DMFT-HF}
and verificatoin of the multiorbital nature of the FS
\cite{comment,Fulde-comment,LDADMFT_multiporbital,Hat-122}.
Also, $P$-induced second superconducting phase of CeCu$_2$Si$_2$
is an important issue
\cite{Hol-122}.


\acknowledgements
We are grateful to P. Fulde, Y. Matsuda, I. Ishida,
T. Shibauchi, Y. Mizukami, S. Kittaka, S. Onari and Y. Yamakawa 
for useful comments and discussions.
This study has been supported by Grants-in-Aid for Scientific
Research from MEXT of Japan.



\clearpage

\makeatletter
\renewcommand{\thefigure}{S\arabic{figure}}
\renewcommand{\theequation}{S\arabic{equation}}
\makeatother
\setcounter{figure}{0}
\setcounter{equation}{0}
\setcounter{page}{1}
\setcounter{section}{1}

\begin{widetext}
\begin{center}
{\bf 
[Supplementary Material] \\
Hexadecapole fluctuation mechanism for $s$-wave 
heavy fermion superconductor CeCu$_2$Si$_2$: \\
Interplay between intra- and inter-orbital Cooper pairs
}%
\end{center}

\begin{center}
Rina Tazai and Hiroshi Kontani
\end{center}

\begin{center}
\textit{Department of Physics, Nagoya University, Nagoya 464-8602, Japan}
\end{center}

\end{widetext}

\appendix

\subsection{A: Model Hamiltonian}
\begin{figure}[htb]
\includegraphics[width=0.4\linewidth]{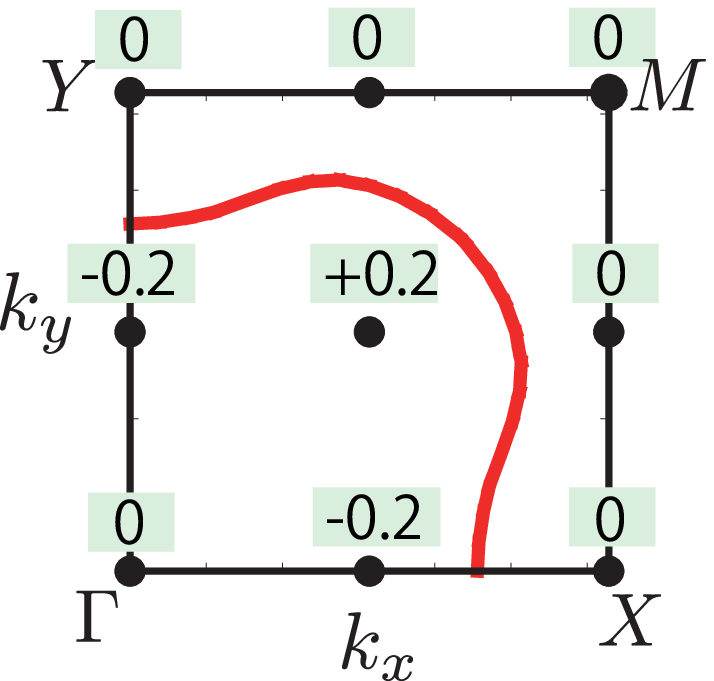}
\caption{The FS with $f$-$f$ hopping.
Each number at $\k$ shows intra-orbital energy shift $\delta E_{\k}$.
}
\label{fig:appendix}
\end{figure}

In this section, we introduce the $f$-$f$ hopping integrals,
by which the $f_{l}$-orbital weight has momentum-dependence on the FS.
The obtained $f$-electron kinetic term is
\cite{S-Tazai-PRB2018}
\begin{eqnarray} 
\hat{H}_{ff}=\sum_{\k l\sigma}E_{\k l}f^{\dagger}_{\k l\sigma}
f_{\k l\sigma}.
\end{eqnarray}
Here, we set $E_{\k1}\equiv E_{1}+\delta E_{\k}$ and $E_{\k2}\equiv E_{2}-\delta E_{\k}$.
To reproduce the $\k$-dependent $\delta E_{\k}$ 
shown in Fig.\ref{fig:appendix},
we introduce from the first to fifth neighbor hopping integrals 
according to Refs.\cite{S-Yamakawa-FeSe,S-Tazai-PRB2018}. 
The obtained momentum-dependence of $f_{l}$-orbital weight on the FS
is shown in Fig. 1 (b) in the main text.

As discussed in Ref. \cite{S-Tazai-PRB2018},
the RPA susceptibility is insensitive to $\delta E_{\k}$ 
since the $f_{l}$-orbital DOS, $D_l(\e)$,
is independent of $\delta E_{\k}$.
In the present study,
we verified that both $\chi$-VC and $U$-VC are also insensitive to 
$\delta E_{\k}$.

In HF systems,
the quadrupole susceptibility remains small within the RPA.
To understand this result, 
we examine the $(Q,Q')$ component of normalized Coulomb interaction:
\begin{eqnarray}
U^{Q,Q'}_0=(\vec{Q})^\dagger \hat{U}^0 \vec{Q}' .
\end{eqnarray}
TABLE \ref{tab:appendix} shows the diagonal component
$U^{Q}_0\equiv U^{Q,Q}_0$.
Since $U^{Q}_0$ for the EM channels
is much smaller than that for the MM channels,
the EM susceptibilities are small within the RPA.
Nonetheless of this fact,
EM susceptibilities strongly develop by considering the AL-VC,
since $X^{\rm AL}$ for the EM channel becomes large
when moderate MM fluctuations exist.

\begin{table}[htb]
  \begin{tabular}{|c|c|c|c|c|c|} \hline
  Q &   C & $O_{20}$  & $H_{0}$ & $H_{z}$ & $O_{yz(zx)}$ 
     \\ \hhline{|=|=|=|=|=|=|}
 $U^{Q}_0$ &   -1.3 & -0.18  & 0.17 &  0.34 & 0.27 
 \\ \hline
   \end{tabular}

  \begin{tabular}{|c|c|c|c|c|c|c|} \hline
  Q &   $J_{z}$ &$T_{z}$ &$D_{z(4)}$ & $J_{x(y)}$& $T_{x(y)}$ & $D_{x(y)}$
     \\ \hhline{|=|=|=|=|=|=|=|}
 $U^{Q}_0$ &   0.56 & 0.44 & 0.55 & 0.49 & 0.49 & 0.50 \\ \hline
   \end{tabular}
\caption{Normalized Coulomb interaction for $Q$-multipole channel, $U^{Q}_0$.
}
\label{tab:appendix}
\end{table}

\subsection{B: Pseudospin representation of multipole operators}
Here, we list the multipole operators $\hat{Q}$
in the present CeCu$_2$Si$_2$ model,
which were already explained in Ref. \cite{S-Tazai-PRB2018}.
The EM (even-rank) operators 
in the $4\times4$ matrix form are expressed as
\begin{eqnarray}A_{1}^{+}&&
\begin{cases}
\hat{C}&=\hat{\sigma}^{0}\hat{\tau}^{0} ,  \nonumber \\
\hat{O}_{20}&=\hat{\sigma}^{0} \left(2.00\hat{\tau}^{0}+3.00\hat{\tau}^{z} \right) ,  \nonumber \\
\hat{H}_{0}&=\hat{\sigma}^{0} \left(-5.73\hat{\tau}^{0}+11.5\hat{\tau}^{z}-12.8\hat{\tau}^{x} \right) ,
\end{cases} \nonumber \\ A_{2}^{+}&&
\begin{cases}
\hat{H}^{z}&=-19.8 \hat{\sigma}^{z} \hat{\tau}^{y}  ,
\end{cases} \nonumber \\ E^{+}&&
\begin{cases}
\hat{O}_{yz}&=-3.87 \hat{\sigma}^{x} \hat{\tau}^{y}  , \nonumber \\
\hat{O}_{zx}&=+3.87 \hat{\sigma}^{y} \hat{\tau}^{y}  . \nonumber \\
\end{cases} \label{eqn:eleO}\\
\end{eqnarray}
The MM (odd-rank) operators are given by
\begin{eqnarray} A_{1}^{-}&&
\begin{cases}
\hat{D}_{4}&=+29.8i \hat{\sigma}^{0} \hat{\tau}^{y}  , \nonumber \\
\end{cases} \nonumber \\ A_{2}^{-}&&
\begin{cases}
\hat{J}^{z}&= \hat{\sigma}^{z} \left(0.50\hat{\tau}^{0}+2.00\hat{\tau}^{z} \right)  ,  \nonumber \\
\hat{T}^{z}&= \hat{\sigma}^{z} \left(9.00\hat{\tau}^{0}-1.50\hat{\tau}^{z} \right)  ,  \nonumber \\
\hat{D}^{z}&= -29.8 \hat{\sigma}^{z} \hat{\tau}^{x}  ,
\end{cases}  \\
E^{-}&&
\begin{cases}
\hat{J}^{x}&= -1.12 \hat{\sigma}^{x} \hat{\tau}^{x}   , \nonumber \\
\hat{J}^{y}&= -1.12 \hat{\sigma}^{y} \hat{\tau}^{x}   ,\nonumber \\
\hat{T}^{x}&=  \hat{\sigma}^{x}  \left(3.75\hat{\tau}^{0}-3.75\hat{\tau}^{z} +5.03\hat{\tau}^{x}\right)  , \nonumber \\
\hat{T}^{y}&=  \hat{\sigma}^{y}  \left(3.75\hat{\tau}^{0}-3.75\hat{\tau}^{z} +5.03\hat{\tau}^{x}\right)  , \nonumber \\
\hat{D}^{x}&=  \hat{\sigma}^{x}  \left( 23.0\hat{\tau}^{0}-6.56\hat{\tau}^{z}-3.14\hat{\tau}^{x}\right)  , \nonumber \\
\hat{D}^{y}&=  \hat{\sigma}^{y}  \left( 23.0\hat{\tau}^{0}-6.56\hat{\tau}^{z}-3.14\hat{\tau}^{x}\right)  ,
\end{cases} \label{eqn:magneO} \\
\end{eqnarray}
where $\hat{\sigma}^{\mu}$ and $\hat{\tau}^{\mu}$($\mu=x,y,z$) are Pauli matrices for
 the pseudo-spin and orbital basis, respectively.
$\hat{\sigma}^{0}$ and $\hat{\tau}^{0}$ are identity matrices.

The row and column of the Hermite matrix $\hat{Q}$ for each operator
is given as $L=(l,\s)$, where $l=1,2$ represents the $f$-orbital
and $\s=\uparrow,\downarrow$ represents the pseudo spin.
In the main text, we also introduce the vector representation
defined as $(\vec{Q})_\a= (\hat{Q})_{L,L'}$, where $\a=(L,L')$.

\subsection{C: Analytic expressions of vertex corrections}
From now on, we introduce the analytic expressions of 
$\chi$-VC \cite{S-Tazai-CeB6} and $U$-VC \cite{S-Tazai-PRB2018}
due to AL diagrams.
First, we discuss the $\chi$-VCs,
whose diagrammatic expressions are shown in Fig. 2 (a) in the main text.
The expression for the AL1 term is given as
\begin{eqnarray}
X^{\rm{AL1}}_{\a \b}(q)=\frac{T}{2}\sum_{\a' \a'' \b' \b'' p}
C_{\a' \b''}^{\a} (q,p) V_{\a' \b'} (p-q) \nonumber \\
\times  V_{\a'' \b''}(p) C_{\b' \a''}^{\b *} (\bar{q},\bar{p}),
\label{eqn:UALc}
\end{eqnarray}
where $p\equiv (\p, \omega_{j})$, $\bar{p}\equiv (\p,-\omega_{j})$, 
and
$\hat{V}(q)\equiv u^{2}\hat{U}^0\hat{\chi}(q)\hat{U}^0+u\hat{U}^0$
is the dressed interaction given by the RPA.
The three-point vertex in Eq. (\ref{eqn:UALc}) is given as
\begin{eqnarray}
C^{EF}_{ABCD} (q,p) \equiv -T\sum_{k}G^{f}_{AF}(k-q)G^{f}_{EC}(k)
G^{f}_{DB}(k-p),
\end{eqnarray}
where $\hat{G}^{f}$ is the $f$-electron Green function.
Also, the expression for the AL2 term is given as
\begin{eqnarray}
X^{\rm{AL2}}_{\a \b}(q)&=&\frac{T}{2}\sum_{\a' \b' \a'' \b'' p}
C_{\a' \b''}^{'\a} (q,p) V_{\b'' \b'} (p-q) \nonumber \\
&\times&  V_{\a'' \a'}(p) \tilde{C}_{\a'' \b'}^{'\b} (q,p),
\label{eqn:S-UALc}
\end{eqnarray}
where 
\begin{eqnarray}
C^{'EF}_{ABCD} (q,p) \equiv -T\sum_{k}G^{f}_{BF}(k-q)G^{f}_{ED}(k)G^{f}_{CA}(k-q+p),
\nonumber \\
\tilde{C}^{'EF}_{ABCD} (q,p) \equiv -T\sum_{k}G^{f}_{AE}(k+q)G^{f}_{FC}(k)G^{f}_{DB}(k+q-p).
\nonumber
\end{eqnarray}
%
%
%
The total $\chi$-VC is given by
$\hat{X}^{\rm AL}=\hat{X}^{AL1}+\hat{X}^{AL2}$,
by subtracting the double counting 
second order diagrams of order $u^2$.

Next, we explain the $U$-VC in the gap equation. 
It is given as
\begin{eqnarray}
( {\hat \Lambda}_{kk'})_{LL'MM'} = \delta_{LM}\delta_{L'M'}+
({\hat L}_{kk'})_{LL'MM'}.
\label{eqn:defALc}
\end{eqnarray}
In the main text, we calculate the AL diagrams for ${\hat L}_{kk'}$.
It is expressed as
\begin{eqnarray}
\!\!(\hat{L}_{kk'})_{LL'MM'}\!\!\!\!\!\!&&=\!\frac{T}{2}\!\!\!\sum_{p,ABCDEF}
B_{ABCDEF}^{MM'} (k-k',p,k') \nonumber \\
&&\times V_{LACD} (k-k'+p) V_{BL'EF} (-p),
\label{eqn:UALc}
\end{eqnarray}
where 
\begin{eqnarray}
&& \hspace{-30 pt}B_{ABCDEF}^{MM'} (q,p,k') =
G^{f}_{AB}(k'-p) \nonumber \\
&&\hspace{30 pt}  \times 
\left\{\! C^{''MM'}_{CDEF} (q,p)+
C^{''MM'}_{EFCD}(q,q+p) \right\}
\label{eqn:Bdef3}
\end{eqnarray}
and
\begin{eqnarray}
{C}^{''AB}_{CDEF} (q,p)\!\equiv 
\!-T\sum_{k'}G^{f}_{CA}(k'+q)G^{f}_{BF}(k')G^{f}_{ED}(k'-p).
\nonumber \\
\end{eqnarray}
%

\subsection{D: Gap equation and retardation effect}

\begin{figure}[htb]
\includegraphics[width=.84\linewidth]{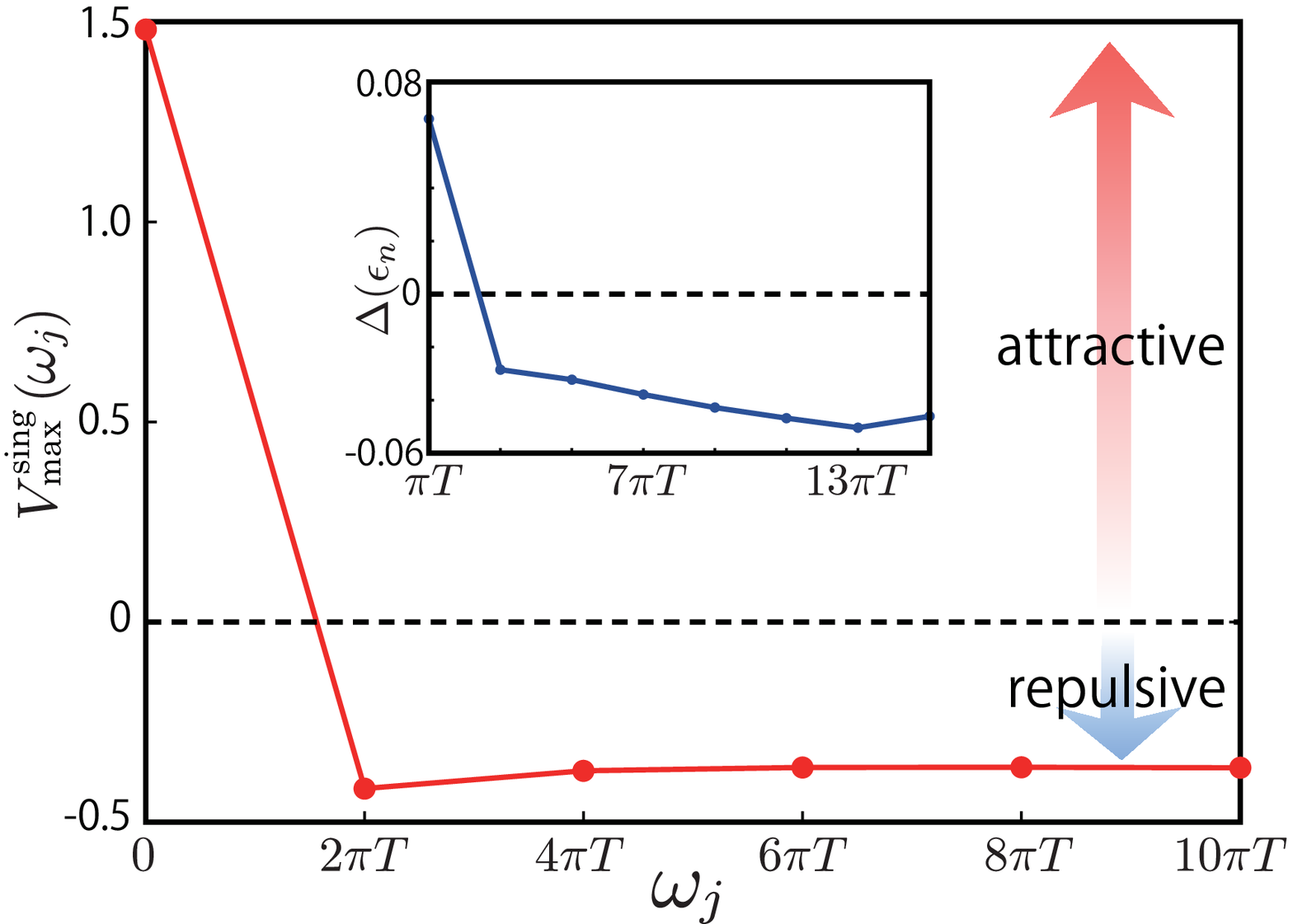}
\caption{
Obtained paring interaction $V_{\rm max}^{\rm sing}(\omega_{j})$ 
and gap function $\Delta(\epsilon_{j})$ (inset) 
as the function of Matsubara frequency. 
Strong retardation effect is recognized.}
\label{fig:chiorb}
\end{figure}

Here, we comment on the retardation effects.
In Fig.\ref{fig:chiorb}, we show the obtained paring interaction 
on the FS defined as
$V^{\rm{sing}}_{\rm max} (\omega_{j})
\equiv \max_{\k,\k'} \{ V^{\rm{sing}}_{(\k,\pi T)(\k',\pi T+\omega_{j})}\}$.
The paring interaction is attractive (positive) at $\omega_{j}=0$,
whereas it becomes to repulsion for $\omega_{j}>0$.
For this reason, the gap function defined as
$\Delta(\epsilon_{n})\equiv \max_{\k} \{ \Delta (\k,\epsilon_{n})\}$
shows the sign-change as the function of $\e_{n}$,
as shown in the inset of Fig.\ref{fig:chiorb}.
This is a hallmark of the retardation effects 
due to the strong $\omega_{j}$-dependence of the 
EM (even-rank) fluctuation.
Since the depairing due to direct Coulomb interaction 
is reduced by the retardation effect,
the fully-gapped $s$-wave superconductivity can be
stabilized in HF systems.


\begin{thebibliography}{99}
\bibitem{comment}
The relation $|V_{\k 1 \sigma}|\approx 2|V_{\k 2 \sigma}|$ holds.
in CeCu$_2$Si$_2$ according to Ref. \cite{LDADMFT_multiporbital}.
Nonetheless, the relation $D_{1}(0)\approx E_{2}(0)$ holds
since $E_1^f$ is higher than $E_f^2$.
In the present model, we set
$|V_{\k 1 \sigma}|=|V_{\k 2 \sigma}|$ and $E_1^f=E_2^f$
to simplify the analysis.
\end{thebibliography}
\end{document}